\documentclass[12pt,a4paper]{article}

\usepackage{a4wide}
\usepackage{amsmath}
\usepackage{bm}
\usepackage{amssymb}
\usepackage{hyperref}
\usepackage{epic}

\newcommand{\DR}{${\mbox{DR}}$ }
\newcommand{\ep}{\epsilon}

\newcommand{\cN}{{\cal N}}

\makeatletter
\def\mr@ignsp#1 {\ifx\:#1\@empty\else #1\expandafter\mr@ignsp\fi}%
\newcommand{\multiref}[1]{\begingroup
\xdef\mr@no@sparg{\expandafter\mr@ignsp#1 \: }%
\def\mr@comma{}%
\@for\mr@refs:=\mr@no@sparg\do{\mr@comma\def\mr@comma{,}\ref{\mr@refs}}%
\endgroup}
\makeatother

\begin{document}

\thispagestyle{empty}


\begin{center}
{\Large{\bf
The Four-Loop Konishi in $\cN=4$ SYM}
}
\vspace{15mm}

{\sc
V.~N.~Velizhanin}\\[5mm]

{\it Theoretical Physics Department\\
Petersburg Nuclear Physics Institute\\
Orlova Roscha, Gatchina\\
188300 St.~Petersburg, Russia}\\[5mm]

\textbf{Abstract}\\[2mm]
\end{center}

\noindent{
We present the result of a \textit{full direct} component
  calculation for the planar four-loop anomalous dimension
of the Konishi operator in $\cN =4$ Supersymmetric Yang-Mills theory.
Our result confirms the results obtained from superfield and superstring
computations, which take into account finite size corrections to the
all-loop asymptotic Bethe ansatz for the integrable models describing the
spectrum of the anomalous dimensions of the gauge-invariant operators and the spectrum of the
string states in the framework of the gauge/string duality.
}
\newpage

\setcounter{page}{1}


The discovery of the AdS/CFT correspondence~\cite{Maldacena:1997re}
with one of its statement
about relation between the spectrum of superstrings on $AdS_5
\times S^5$ and the spectrum of the anomalous dimensions of the gauge-invariant operators in
maximally extended $\cN=4$ Supersymmetric Yang-Mills (SYM) gauge theory
has renewed interest to the perturbative calculations in $\cN=4$ SYM model~\cite{Gliozzi:1976qd}.
First such one-loop calculations\footnote{The anomalous dimension of the
Konishi multiplet has been computed at one (level $g^2$) and two (level $g^4$)
loops through OPE analysis of the four-point function of stress-tensor
multiplets~\cite{Anselmi:1996dd}.} were performed for the quasipartonic twist-2 operators
with an arbitrary Lorentz spin~\cite{LN4} as a generalization
of the famous QCD results~\cite{GWGPAP}.
It was found, that eigenvalues of anomalous dimension matrix
can be expressed through $\Psi$-function with shifted argument~\cite{LN4}.
This result allowed author to make a suggestion,
based on the study of integrability~\cite{Lipatov:1993yb}
related with Balitsky-Fadin-Kuraev-Lipatov (BFKL) equation~\cite{Lipatov:1976zz},
that the computations of the anomalous dimensions of Wilson operators in $\cN=4$ SYM theory should be also
related with integrability.
The integrability was
found by mapping the planar one-loop dilatation operator of $\cN=4$ SYM theory into the Hamiltonian
of a Heisenberg spin chain~\cite{Minahan:2002ve}. 
Generalization on the higher-loop orders has allowed to write all-loop asymptotic Bethe
ansatz~\cite{Beisert:2003yb},
which after adding a dressing factor\footnote{Necessity of a modification of the
asymptotic Bethe ansatz was obtained by direct
calculation of the four-loop MHV-amplitude~\cite{Bern:2006ew}, what
allowed to find the large spin limit of the four-loop anomalous dimension.
The first nontrivial coefficient for the dressing factor
at weak coupling was checked by a direct perturbative evaluation~\cite{Beisert:2007hz}.}
\cite{Beisert:2006ez} stands in the agreement with the Bethe ansatz for the sigma-model from the string
side~\cite{Bena:2003wd}.

Some perturbative calculations were performed, which confirmed the results obtained from the integrability.
The direct calculation of the two-loop anomalous dimensions matrix
for the twist-2 operators with the arbitrary Lorentz spin~\cite{Kotikov:2003fb}
allowed to confirm the
{\it maximal transcendentality} principle for the matrix eigenvalues~\cite{Kotikov:2002ab},
which should have the harmonic
sums
entering into the answer
in the given order of perturbative theory.
Namely,
in the $n^{th}$ order of the perturbative theory the sum
of modules of indices in the harmonic sums is equal to $2n\!-\!1$.
Have this principle in mind one can extract
the anomalous dimension for the twist-2 operators with arbitrary Lorentz spin
in $\cN=4$ SYM theory from
the known result in Quantum Chromodynamics (QCD).
When the results for the three-loop anomalous dimensions in QCD~\cite{Moch:2004pa}
became available after long time calculations,
we easily extracted corresponding result for the three-loop anomalous dimension in $\cN=4$
SYM theory~\cite{Kotikov:2004er}.
This result was found in a perfect agreement with the results from the asymptotic
Bethe ansatz not only for Konishi,
but also for twist-2 operators with higher Lorentz spins.
The result for the three-loop Konishi was then confirmed
by direct calculation~\cite{Eden:2004ua}.

However the asymptotic Bethe ansatz (ABA)
can be applied only for the long-range operators but for the short operators, such as Konishi,
it will be breakdown by so called ``wrapping'' effect. Indeed, it was shown~\cite{Kotikov:2007cy},
that the ABA
give at the four loops the result, which is in the contradiction with the predictions coming from the BFKL
equation.
To take into account ``wrapping'' effect one should make some additional calculations
either in the perturbative theory or in the superstring theory.
The first one were performed in Ref.~\cite{Fiamberti:2007rj},
where the special
class of four-loop diagrams were calculated in the superfield formalism.
Superstring calculations were made in Ref.~\cite{Bajnok:2008bm}
by taking into account so called L\"uscher-terms~\cite{Luscher:1986pf}.
Both results were obtained for Konishi and now are in the agreement
after corrections from the perturbative calculations side.
Because both calculations have some of suggestions
and use the result obtained from the ABA,
it will be nice to have as a check
the same result from the full direct calculations of the four-loop anomalous dimension of the Konishi
operator.
In this
paper we have presented the result of the {\it full} calculation for
the planar four-loop anomalous dimension of the Konishi operator
in $\cN=4$ SYM theory directly in the components, using Laporta
algorithm~\cite{Laporta:2001dd}, which
based on the resolution of the integration by part (IBP) identities and
was
successfully applied for the calculation of the four-loop renormalization constants in
QCD~\cite{Czakon:2004bu},
with the method from Refs.~\cite{Misiak:1994zw}.


The Konishi operator is the most simple operator in the $\cN=4$ SYM theory and is nothing then
the kinetic term of the chiral matter superfields~\cite{Konishi:1983hf}:
\begin{equation}
\mathcal{O}_\mathcal{K}=\mathrm{tr}\ e^{g_{{}_{YM}}V}\bar\Phi_I e^{-g_{{}_{YM}}V} \Phi^I\,,\qquad g=\frac{\sqrt{\lambda}}{4\pi}\,,
\end{equation}
where $\lambda=g^2_{YM}N_c$ is the t'Hooft coupling constant.
Its low lying state in components is:
\begin{equation}
\mathcal{O}_\mathcal{K}=\mathrm{tr}\ \bar\phi_{i}\phi^{i}=\mathrm{tr}\left[A_{i}A^{i}+B_{i}B^{i}\right],\qquad i=1,2,3\,,
\end{equation}
where $\phi^i$ is the complex adjoint scalar field, while $A^i$ and $B^i$ are the real adjoint
scalar and pseudoscalar
fields correspondingly.

The calculation of the anomalous dimension of such operator in an usual way~\cite{GWGPAP} will
rise the infrared divergencies (see Refs.~\cite{Gracey:2002yt} about the same
problem for the similar operator $A^\mu A_\mu$).
To avoid its one should allow to flow a momentum through the operator and calculate the vertex type diagrams
instead of the initial propagator type diagrams. However we can nullifying one of the momenta flowing through
one of the external scalar field without appearance of any IR divergencies. This trick,
based on the
Infrared Rearrangement~\cite{Vladimirov:1979zm}
or in more general on the
$\mbox{R}^*$-operation~\cite{Chetyrkin:1982nn}\footnote{See Ref.~\cite{Czakon:2004bu} for details.},
is used
in the calculation of the renormalization constants of the vertices and its application for the
calculation of the anomalous dimension of the similar operator $A^\mu A_\mu$
 can be found in
Refs.~\cite{Gracey:2002yt}.
After nullifying one of the external momenta we effectively obtain again the propagator type diagrams,
which can be easily evaluated
by FORM~\cite{Vermaseren:2000nd} package MINCER~\cite{Gorishnii:1989gt}
up to three loops with the method from
Refs.~\cite{Larin:1993tp}.
However for the four-loop calculations MINCER is still
unavailable but we can use the method of Laporta~\cite{Laporta:2001dd}.
To use this method for our four-loop calculations we following
Refs.~\cite{Misiak:1994zw} expand all propagators on the external momentum to make
the divergence of the diagram logarithmic, then nullifying external
momentum in propagators (only in denominators of corresponding Feynman
integrals, not in numerators) and make all lines
massive with the same mass. In this way we obtain massive tadpole diagrams
instead of initial massless propagator type diagrams. In the case of our
calculations all diagrams have already logarithmic divergence and we do not
need expansion at all. For obtained tadpoles we apply the method of Laporta~\cite{Laporta:2001dd},
which is based on the resolution
of the IBP identities and has some of implementations in the different languages/CAS such as
AIR~\cite{Anastasiou:2004vj}, DiaGen/IdSolver~\cite{idsolver} and FIRE~\cite{Smirnov:2008iw}.
We have used
our own implementation of the Laporta's algorithm
\cite{Laporta:2001dd} in the form of the MATHEMATICA package BAMBA~\cite{BAMBA} with the master integrals from
Ref.~\cite{Czakon:2004bu}. Our realization is very close to AIR~\cite{Anastasiou:2004vj} with
some improvements, major from which is the usage of the symmetry of the tadpole
integrals\footnote{Results for integrals can be obtained under request.}.

For the dealing with a huge number of diagrams we have used a program DIANA~\cite{Tentyukov:1999is},
which call QGRAF~\cite{Nogueira:1991ex} to generate all diagrams.
We have written few routines, which allowed us considerable simplified a work with the topologies.

As important check of our program we reproduced the part of the anomalous dimension of the gluon
field~\cite{Czakon:2004bu}
coming from the pure Yang-Mills gauge theory because the diagrams, which give contributions to this part,
have all possible topologies in the four-loop order for the propagator type diagrams.

The calculations were performed with dimensional
reduction ($\DR$) scheme~\cite{Siegel:1979wq} and the Feynman rules from
Refs.~\cite{Gliozzi:1976qd,Avdeev:1980bh},
which were used for the calculation
of the $\beta$-function
in the $\cN=4$ SYM theory up to three loops~\cite{Avdeev:1980bh}.
In principal, $\DR$-scheme should be violated
in higher-loop orders~\cite{Siegel:1980qs},
but we have found\footnote{A fact, that the result of Refs.~\cite{Avdeev:1982np} is incorrect
was pointed out firstly in Ref.~\cite{Harlander:2006xq}.}
\cite{rc}, that at least up to three loops the $\DR$-scheme works correctly for
the calculations of the renormalization constants of the triple vertices
on the contrary of the result from Refs.~\cite{Avdeev:1982np}. This our result~\cite{rc}
allows to hope, that the $\DR$-scheme should work in the four loops at least for the calculations of
the propagator type diagrams.

A total number of four-loop diagrams is 131015.
All calculations were performed in the Feynman gauge with FORM~\cite{Vermaseren:2000nd},
using FORM package COLOR~\cite{vanRitbergen:1998pn} for evaluation of the color traces.
For the renormalization we have used the renormalization constants from Ref.~\cite{rc}.
In addition, we need the counterterms for the gluon and scalar ``masses'':
\begin{eqnarray}
%
Z_{m_{g}}^{3-loop}&=& 
\frac{36}{\ep}\,g^2
+\left(-\frac{603}{4\;\ep^2} + \frac{111}{4\;\ep}\right)\,g^4
\nonumber\\
&&+\left(\frac{977}{4\;\ep^3} + \frac{27335}{48\;\ep^2}
+\frac{1}{\ep}\left[
\frac{43745}{96} - \frac{1539}{2}\,S_2 -
\frac{651}{8}\,\zeta(2) - \frac{217}{2}\,\zeta(3)\right]\right)\,g^6\,,\\
Z_{m_{\phi}}^{3-loop} & = & 
-\frac{36}{\ep}\,g^2
+\left(\frac{60}{\ep^2} + \frac{387}{2\;\ep}\right)\,g^4
\nonumber\\
&&+\left(\frac{5155}{12\;\ep^3} - \frac{4603}{12\;\ep^2}
+\frac{1}{\ep}\left[\frac{11543}{4} + 504\;S_2 - \frac{231}{4}\,\zeta(2)
- 235\,\zeta(3)\right]\right)\,g^6\,,
\end{eqnarray}
with $S_2$ from Ref.~\cite{Czakon:2004bu}.
Origin of these counterterms
is rather simple. When one calculate the inverse gluon or scalar propagators
with the method, which we have used, the
result consists of two terms. The first term is proportional to the square of the external
momentum and this is the renormalization constant of the gluon (or scalar) field. The second
term is proportional to the mass, which was introduced for the IR-regularization of
the scalar integrals. This means, that when we will make the renormalization
in the next orders of the perturbative theory we should subtract
exactly the same two structures both for the gluon and scalar propagators.

The final result
after subtraction of the anomalous dimension for the scalar field is:
\begin{equation}\label{resk}
\gamma^{\mathrm{4-loop}}_{\mathcal{K}}\ =\ 12\,g^2
-48\,g^4+336\,g^6+\Big(-2496+576\,\zeta(3)-1440\,\zeta(5)\Big)\,g^8
\end{equation}
in a full agreement with the results of Refs.~\cite{Fiamberti:2007rj}, \cite{Bajnok:2008bm}.

An important check of our result (\ref{resk}) is the absence of some special numbers such as
$\zeta(2)$, $\zeta(4)$, $S_2$ and other, which enter in the scalar master integrals
from Ref.~\cite{Czakon:2004bu}.

As a by product we have obtained the following result for the four-loop
anomalous dimension for the scalar field in the planar limit:
\begin{equation}
\gamma^{\mathrm{4-loop}}_{\phi}=4g^2
-2g^4+\left(\frac{23}{2}-27\,\zeta(3)\right)g^6+
\left(\frac{1669}{24}+\frac{423}{4}\zeta(3)+\frac{57}{4}\zeta(4)-290\zeta(5)\right)g^8.
\end{equation}

Our result of the \textit{full direct} component calculation for the planar four-loop anomalous dimension
of the Konishi operator in $\cN =4$ SYM theory (\ref{resk}) confirms that indeed the additional contribution to the
anomalous dimension of the operators coming from the ``wrapping'' effect can be
calculated both from the perturbative theory~\cite{Fiamberti:2007rj}
and from the superstring theory~\cite{Bajnok:2008bm} by taking into account
the corrections to ABA from the additional diagrams or from the L\"uscher-terms correspondingly.
We want to stress that our calculations don't contain any suggestions at all and can
be considered as unique ``experimental'' check for the computations from
Refs.~\cite{Fiamberti:2007rj,Bajnok:2008bm} including the correctness of the
asymptotic Bethe ansatz for the planar AdS/CFT system.

 \subsection*{Acknowledgments}
We would like to thank Lev Lipatov, Andrei Onishchenko and Matthias Staudacher for useful discussions.
This work is supported by
RFBR grants 07-02-00902-a, RSGSS-3628.2008.2.
We thank the Max Planck Institute for Gravitational Physics
at Potsdam for hospitality while working on parts of this
project.

\end{document}